\documentstyle[titlepage]{article}
\def\ref{\par\noindent\hangindent=1cm}
\begin{document}
\newcommand{\N}{$^{1,2}$}
\newcommand{\M}{$^{3}$}
\def\mincir{\raise -2.truept \hbox{\rlap{\hbox{$\sim$}}\raise 5.truept
\hbox{$<$}\ }}
\def\magcir{\raise -2.truept \hbox{\rlap{\hbox{$\sim$}}\raise 5.truept
\hbox{$>$}\ }}

\begin{titlepage}
\begin{center}
\vspace{1cm}
\Huge
{The Local Galaxy Density and the \\ Bulge-to-Disk Ratio of Disk Galaxies}
\large
\vspace{2cm}

Giuliano Giuricin\N, F\"usun Limboz Tektunali\M, Pierluigi Monaco\N, \\
Fabio Mardirossian\N, Marino Mezzetti\N \\

\end{center}
\normalsize
\bigskip
\bigskip

1) Scuola Internazionale Superiore di Studi Avanzati (SISSA),
via Beirut 4, 34013 --- Trieste, Italy.

2) Dipartimento di Astronomia, Universit\`a di Trieste,
via Tiepolo 11, 34131 --- Trieste, Italy.

3) Istanbul University Observatory, 34452 Beyazit --- Istanbul, Turkey.

\bigskip

\noindent email:\\
giuricin, mardirossian, mezzetti, monaco@tsmi19.sissa.it\\
astro02@triuvm11.bitnet

\vspace{2.5cm}

\begin{center}
\section*{Abstract}
\end{center}

Relying on samples of disk galaxies for which a detailed photometric bulge/disk
decomposition has been provided in the literature, we examine the dependence of
the bulge-to-disk luminosity ratio (B/D) on the blue absolute luminosity and
on the environmental density. In our statistical analysis of various B/D data
sets we pay particular attention to disentangling the role played by the galaxy
morphology--galaxy density relation. Besides, we focus our attention on nearby
($z<0.01$) galaxies, for which we can provide a three-dimensional
characterization of the local galaxy density.

We find that the observed tendency of galaxies to have greater B/D with
increasing galaxy density simply reflects the average decline of B/D towards
later morphological types together with the morphology--density relation. This
relation tends to give rise also to a greater proportion of bright bulges in
denser regions, because the decrease of B/D towards later types is mostly due
to a dimming of the bulge rather than to a brightening of the disk. But when
we remove the effect induced by the morphology--density relation, we detect no
clear evidence of a dependence of B/D on galaxy density.
Furthermore, B/D turns out to be substantially unrelated to the blue absolute
magnitude of the galaxy. We briefly discuss to what extent our results
(partially) disagree with previous claims.

\bigskip

{\it Subject headings:} galaxies: clustering --- galaxies: structure

\bigskip
\bigskip
\begin{center} SISSA Ref.: 28/95/A \end{center}
\end{titlepage}


\section{Introduction}

A major task of the theories of galaxy formation and evolution is to account
for the distinct components into which the galaxy light distribution can be
divided, the most prominent components being the bulge and the disk. Much
work has been devoted to the photometric separation of bulge and disk light
contributions and to the study of the main structural properties of these
components (see, e.g., the reviews by Simien, 1988, and by Capaccioli \& Caon,
1992). However, insufficient attention has been paid to exploring the influence
of the galactic environment on these properties, in particular on the
bulge-to-disk luminosity ratio (hereafter B/D); this quantity tends to
decrease towards later morphological types as a natural consequence of the
Hubble classification criteria. The Hubble morphological sequence from
early to late types is largely a sequence of increasing dominance of disk
over bulge. But the weakness of the dependence of B/D on morphological
type (see  \S 2 below) in practice makes it impossible to view the Hubble
morphological sequence simply as a B/D sequence only.

{}From an analysis of his own estimates (Dressler, 1980a) of
the total and bulge magnitudes of a very large sample of cluster galaxies,
Dressler (1980b) noted that, for both lenticulars and spirals, B/D and bulge
luminosity (though not total luminosity) show a positive correlation with the
local projected galaxy density (computed for each galaxy on the basis of the
area encompassing the ten nearest projected neighbors). Considering also a
nearly complete magnitude-limited sample of field galaxies, Schechter \&
Dressler (1987) confirmed that cluster galaxies have greater B/D than field
galaxies; furthermore, they reported a strong positive dependence of B/D on
the galaxy blue total luminosity for generic samples of cluster and field
galaxies; this relation was found to be weaker for spirals alone.

Later, Solanes, Salvador-Sol\'e \& Sanrom\`a (1989) reanalyzed Dressler's
(1980a) data, subdividing disk galaxies into four morphological types (S0,
Sa, Sb, Sc). They noted a significant B/D--absolute magnitude correlation,
especially for late types (Sb and Sc), with brighter galaxies having an average
lower B/D (i.e., contrary to the claim by Schechter \& Dressler
(1987)). Besides, they emphasized that the claimed environmental density
effects on B/D simply reflect the well-known morphology--local density relation
(see, e.g., Whitmore, Gilmore \& Jones, 1993, and Whitmore, 1990), which holds
true also within spirals (e.g., Giuricin et al., 1988; Tully, 1988b), owing to
the average decline of B/D towards later morphological types. In other words,
the fact that earlier types, which have greater B/D, preferentially reside in
denser regions would create a B/D--environmental density correlation as a
secondary effect, which would vanish within subsets of objects of a given
morphological type.

In the present paper we wish to address all these issues, viz. the relations of
B/D to absolute magnitude and to environmental density at both small and large
scales (large-scale effects have not yet been investigated in the literature).
For this purpose we shall rely on the galaxy samples with available bulge/disk
decomposition provided by Kent (1985, 1986, 1988; hereafter K85, K86, K88),
Simien \& de Vaucouleurs (1986), (SdV), Kodaira, Watanabe \& Okamura (1987)
(KWO). These samples, albeit much smaller than Dressler's (1980a), are
characterized by a detailed bulge/disk decomposition, a much finer and more
reliable morphological classification, a more accurate galaxy apparent optical
magnitude, and a better knowledge of galaxy environment and distance (the
redshift is known for all galaxies, whilst redshift incompleteness is large
in Dressler's (1980a) sample). They comprise mostly bright and nearby galaxies,
which span a rather wide range of environmental density (from the Virgo cluster
to the low-density regions of the Local Supercluster). We have also taken into
consideration the data sample provided by Andredakis \& Sanders (1994) (AS),
which, however, contains a small number of objects.

In \S 2 we describe the data samples used and our parameters for
local galaxy density. In \S 3 we present our analysis of the
B/D data. The conclusions are reported in \S 4.

\section{The Data Samples}

\subsection{The B/D data samples}

Most of the classical methods of bulge--disk photometric decomposition rely on
fundamental differences in the light profiles or the intrinsic flattenings
of the two components. Adopting standard laws for the light distribution of the
bulge and disk components (i.e., a de Vaucouleurs $r^{1/4}$ and an exponential
law, respectively), K85 employed mostly a standard non-linear least-squares fit
to the main axis profiles of 105 disk galaxies. For a few objects he used the
iterative decomposition method described by Kormendy (1977), in which one
solves for the disk parameters in a region where disk light dominates, and
likewise for the bulge parameters. At each iteration, the light from the
component being kept fixed is subtracted from the total luminosity profile
before the other component is solved for. The process is iterated until
convergence is achieved. In his subsequent photometric study of 37 Sb and Sc
objects (K86), without making any assumption regarding the fitting laws for
either component, the author assumed that each one is characterized by
elliptical
isophotes of constant, and essentially different, flattening; then, an
iterative process yielded the bulge and disk profiles. Later, following one of
the two approaches used in his previous studies, Kent (1988) reported the
luminosity--profile decomposition of another 14 spirals. Kent (1985, 1986,
1988)
reduced his photometric measures to the $r$ band of the $uvgr$ system of Thuan
\& Gunn (1976). We have also assembled together the K85, K86, K88 data in
order to obtain the fairly homogeneous global Kent (K) sample.

SdV slightly modified Kormendy's iterative decomposition procedure in order to
allow a better handling of individual luminosity profiles. Fitting the
above-mentioned standard laws to the light profiles, SdV provided new results
for 32 galaxies and collected results on 66 additional objects from previous
papers. They homogeneized the merged set of B/D in the $B$ band by making
some corrections to the original B/D values in a few cases for which the
results of various authors showed systematic differences.

Without resorting to an initial guess for the bulge and disk parameters, KWO
generated a grid of models with different combinations of scale parameters;
from among these they selected the ten best-approximating models, consisting of
a standard spheroid and a standard disk. Later, they made a search among these
models to identify the one giving the best fit to the observed profile. KWO
presented structural parameters (in the V band) of 167 galaxies which are
probable members of the Virgo cluster and the Ursa Major cloud. Their galaxy
sample is nearly complete for large and bright galaxies ($\log D_{26}$(0.1')
$\magcir$1.3 and $V_{26}\mincir$14 $mag$) in these two sky areas.

Furthermore, AS reanalyzed the K86 observational data by means of different
decomposition schemes; they applied a standard non-linear
two-component (bulge and disk) simultaneous fitting to the
observed light profile. Besides modeling spirals by an exponential disk
and a $r^{1/4}$-law bulge, the authors also attempted to decompose spirals in
terms of an alternative model, consisting of an exponential disk and an
exponential bulge; they concluded that the double-exponential model is
superior in many statistical aspects of the fitting procedure for the
bulges of mid- and late-type spirals. The two sets of B/D values which result
from the two approaches are hereafter referred to as AS1 and AS2, respectively.

In the present study we consider the lenticulars and spirals of the galaxy
samples mentioned above. More specifically, we shall mainly analyze the
homogeneous samples SdV, KWO, K. Omitting from the KWO sample the galaxies
with low-quality
spheroids and disks (i.e., with respective weights $w=2$ and $w<4$, as
tabulated by KWO in their Table 2), we also construct a
homogeneous subsample of good quality data (hereafter referred to as KWO'
sample). We also examine samples K85, K86, K88, AS1, AS2, which, however,
include a small number of objects.  In the case of multiple entries for a
galaxy, we adopt the mean values of B/D. In the statistical analysis described
in the following section, more confidence should be given to the the spirals of
the K sample than to the whole S0+S K sample, because K85 regarded his
own decompositions resulting in B/D$>$1.70 as particularly dubious (objects
with such a great B/D are generally ellipticals or lenticulars). The B/D data
given by SdV and K85 are corrected for internal absorption; the other data sets
are uncorrected, but the authors generally avoided observing strongly inclined
objects.

\subsection{Comparison between different samples}

The comparison between different data samples is not an easy task, because of
the inhomogeneity of photometric measures, which are taken in different
spectral bands, and even more because of the different photometric
decomposition procedures (although some attempts to merge different samples
will be made in the following section). The inhomogeneity of different data
samples is illustrated in Table 1 and Fig. 1. Table 1 reports the means,
standard deviations s.d., medians, associated 90\% confidence intervals of the
B/D distributions of samples SdV, K, KWO, KWO', AS1, AS2 for various
morphological types. Fig. 1 shows the plot of the medians of B/D (and of
associated 90\% confidence intervals) versus morphological type, for
samples SdV, K, KWO, KWO'. We have verified that colour differences alone
cannot account for the differences in the B/D data sets; thus, the observed
basic differences between different B/D samples are to be mainly ascribed to
systematic effects in photometric decomposition techniques.

In the following lines we briefly comment on the most pronounced differences
between different B/D samples (see, e.g., Simien (1988) for further
details). The B/D distributions (as well as structural parameters) derived
by K and SdV are in broad agreement for lenticulars and early-type spirals;
for late types, SdV found typically greater B/D values than K did. For
late-type spirals, KWO found bulges of systematically larger effective radii
and luminosities, but lower effective brightnesses than SdV and K85 did (see,
e.g., Simien, 1988). As a consequence, their (typically greater) B/D values
show a weaker tendency to decrease along the Hubble sequence with respect to
the SdV and K data. However, the KWO' subset of good quality data shows better
agreement with the SdV and K data.  With their new double-exponential model,
AS2 drastically reduced the range of values spanned by the structural bulge
parameters (effective radius and surface brightness). The smaller scale lengths
and fainter effective surface brightnesses which result from their alternative
bulge model also led to smaller bulge luminosities and B/D values. For mid- and
late-type spirals, the AS2 B/D data are in closer agreement with K86 than the
AS1 data are (whilst for early-type spirals the AS2 data seem unreasonably
low.)

In general, B/D, bulge and disk luminosities are thought to be more reliably
determined than the structural parameters of the two components (e.g.,
Schombert \& Bothun, 1987), which are therefore not analyzed in the present
paper. Remarkably, in all the samples considered, the typical decrease of B/D
towards later types is due much more to a drop in bulge luminosity than to a
brightening of the disk (as was in general already noted by the authors of
the photometric decompositions). This important point will be discussed in
the last section.

There are 12 galaxies (9 spirals) common to the SdV and K samples. The standard
deviations of the differences between the B/D values of the galaxies in common
are s.d.=2.57 and 0.54, for the 12 disk galaxies and 9 spirals,
respectively. These observed standard deviations can be compared with the
respective values of s.d.=2.02 and 1.38, which are the standard deviations
which would be expected if the whole scatter in B/D (as represented by the
standard deviations tabulated in Table 1) were due to errors in the
determination of B/D. We evaluate the expected standard deviations as the
square roots of the weighted mean variances, taking weights equal to the
proportions of common objects of various morphological types. An analogous
exercise for the 30 spirals common to K86 and AS1 (K86 and AS2), who analyzed
the same photometric data in different manners, leads to the observed value of
s.d.=1.00 (s.d.=0.51) --- which is due to differences in decomposition
only --- as against the expected value of s.d.=1.18 (s.d.=0.66). For
the 19 objects (9 spirals) common to the SdV and KWO samples, we obtain
observed
values (s.d.=1.28 and s.d.=0.19) which are definitely smaller than the
expected values (s.d.=1.90 and s.d.=0.77). The same holds true for 11
objects (7 spirals) common to the SdV and KWO' samples, for which we obtain the
observed values of s.d.=1.24 and s.d.=0.18 as against the expected
values of s.d.=1.57 and s.d.=0.56.

We conclude that most of the dispersion in B/D is due to errors in
decomposition (with slight contributions from errors in photometry and
morphology classifications). But there is certainly some cosmic scatter in B/D
(for spirals in all samples and for lenticulars in at least some samples).
Thus, we may well wonder whether environmental density can be a major source
of intrinsic scatter in B/D.

\subsection{The environmental density}

As already done in recent studies on the environmental effects on bars
(Giuricin et al., 1993), arm classes (Giuricin et al., 1994), and nuclear
activity (Monaco et al., 1994), we use the Nearby Galaxy Catalog (Tully, 1988a;
NBG) to give a three-dimensional definition of the environment. This catalogue
is intended to include all known nearby galaxies with systemic velocities lower
than 3000 km s$^{-1}$, which corresponds to a distance of 40 Mpc with the
Hubble constant $H_0=75$ km s$^{-1}$ Mpc$^{-1}$ (the value adopted throughout
the present paper). In the NBG every galaxy is assigned a distance based on its
redshift, on the assumed value of $H_0$, given before, and on corrections for
group membership and Virgo infall (according to the Virgocentric retardation
model described by Tully \& Shaya, 1984). Every galaxy member of a group or
cluster is given a distance consistent with the mean velocity of the system
itself.

Following Tully (1988b) in the main, with this spatial distribution of galaxies
we define the parameter $\rho_{\sigma}$ of local galaxy density (where
$\rho_{\sigma}$ is in units of galaxies per Mpc$^{3}$) as the number of
galaxies per Mpc$^{3}$ that are found around every galaxy within the smoothing
length $\sigma$ (in Mpc): \begin{equation}
\rho_{\sigma}=\sum_i C \exp[-r_i/2\sigma^{2}]
\end{equation} where every galaxy is smoothed with a gaussian filter of
half-width $\sigma$, $r_i$ is the spatial distance of the $i$-th galaxy from
the specified galaxy and the normalization coefficient is $C=1/(2 \pi
\sigma)^{3/2}=0.0635/\sigma^{3}$; the sum is carried out for all galaxies
except
the one we are calculating the density for. In order to correct the density for
incompleteness of NBG at large distances, we weight every galaxy with
a correction factor $F(\mu)$, where $\mu=5 \log D + 25 $ is the distance
modulus and $D$ is the distance in Mpc. $F$ expresses the number of galaxies
(brighter than $M_B=-16$) that exist for every galaxy catalogue. We use the
following expression for $F$:
\begin{equation} F = \exp [0.033 (\mu-28.5)^{2.83}]
\end{equation} (and $F$=1 when $\mu<28.5$). Finally, our density parameter,
corrected for incompleteness, is
\begin{equation} \rho_{\sigma} = \sum_i C F(\mu_i) \exp [-r_i^{2}/2\sigma^{2}];
\end{equation}
$\rho_{\sigma}$ gives the number of galaxies, per Mpc$^{3}$,
brighter than $M_B=-16$ around the galaxy considered (see Monaco et al., 1994,
for the evaluation of $F(\mu)$ and $\rho_{\sigma}$).

As discussed in Giuricin et al. (1993), because of the clustering properties of
galaxies, the choice of different $\sigma$-values implies a different physical
meaning for the local galaxy density $\rho_{\sigma}$, so that a dependence of a
quantity on $\rho_{\sigma}$ with low (high) $\sigma$-values refers to small-
(large-) scale density effects. Border effects make less reliable the estimates
of $\rho_{\sigma}$ for objects which lie close to the limiting distance of the
NBG sample; however, there are only a few NBG galaxies lying at distances
greater than 36 Mpc in our samples (i.e., one, six, and two objects in the SdV,
 K, KWO and KWO') samples, respectively). In any case, we have checked that the
omission of these few objects does not substantially change the results
presented below (\S 3).

In the present paper we have updated the galaxy morphological types by
consulting the Third Reference Catalogue of Bright Galaxies by de
Vaucouleurs et al. (1991) (RC3).

\section{ Analysis and Results}

First, we analyze separately the lenticulars and spirals (S0+S) of the
homogeneous samples SdV, KWO, KWO', K. The analysis of samples
K85, K86, K88, AS1, AS2 does not yield interesting results, because of poor
statistics. We shall also attempt to
merge different B/D samples. Basically, we consider those objects (lenticulars
and spirals) which are included in NBG and we repeat the analyses for the
subsets of spirals (S) and spirals having morphological types from T=0 to T=5;
this is roughly the T range where usual photometric decomposition techniques
are thought to be more accurate, since both bulge and disk contributions are
important (see, e.g., the methodological discussion by Schombert \& Bothun,
1987).

In the following discussion of our correlation analysis, we shall present only
the most interesting results in some tables and we shall mention the
cases for which we obtain significance correlations. In the discussion we shall
not bother with correlations at the (one-tailed) $<$95\% significance level
(which corresponds to the $<$90\% level for a two-tailed test).

\subsection{The B/D--$\rho_{\sigma}$ and T--$\rho_{\sigma}$ correlations for
homogeneous samples}

In order to investigate whether environmental density influences B/D, we
primarily deal with the B/D--$\rho_{\sigma}$ correlations (with $\sigma$=0.25,
0.5, 1, 2 Mpc) taking into account the correlation of B/D with galaxy
morphological type T (coded as in RC3). We analyze the significance of the
correlations between two variables by computing the two non-parametric rank
correlation coefficients, Spearman's $r_s$ and Kendall's $r_k$ (see, e.g.,
Kendall \& Stuart, 1977). In order to estimate to what extent the
B/D--$\rho_{\sigma}$ correlations are spuriously induced by the
morphology--density (T--$\rho_{\sigma}$) relations, whenever the latter
correlations are statistically significant, we calculate Kendall's partial
correlation coefficient $r$ (e.g., Siegel, 1956). This is a measure of the
correlation between two data sets (B/D and $\rho_{\sigma}$, in our case)
independently of their correlation with a third data set (T, in our case).
Since the sampling distribution of $r$ is unknown, we adopt the bootstrap
method of resamplings (e.g., Efron, 1979; Efron \& Tibshirani, 1985) in order
to compute its statistical significance (we performed 5000 bootstrapping
resamplings for each correlation).  We calculate the partial correlation
coefficient $r$ and its statistical significance when the
T--$\rho_{\sigma}$ correlation is significant (at the $>$90\% confidence
level).

In Table 2 we present the results of the most interesting cases; we list the
two correlation coefficients $r_s$ and $r_k$, together with the associated
(one-tailed) percent significant levels for the correlations of B/D versus
$\rho_{\sigma}$ (with $\sigma$=0.5 Mpc) and of T versus $\rho_{\sigma}$ (with
$\sigma$=0.5 Mpc). The correlations involving $\rho_{\sigma}$ with
$\sigma$=0.25, 1, and 2 Mpc lead to similar results.

We note that the expected T--$\rho_{\sigma}$ correlations turn out to be very
significant within wide samples (SdV, KWO, KWO') of S0+S objects (they are
perhaps present also within the SdV spirals). These correlations indicate that
earlier types preferentially reside in denser regions than later types. Poor
statistics probably prevent the appearance of the morphology--density relation
in the other, generally smaller, samples. (For the K sample there
are few objects which reside in high-density regions.)  Also the
B/D--$\rho_{\sigma}$ correlation is very strong within wide samples (SdV, KWO,
KWO') of S0+S objects and marginal within KWO' spirals. Figs. 2 and 3 show
the B/D--$\rho_{0.5}$ correlations for the S0+S objects of the SdV and KWO
samples. Since various morphological type intervals are denoted by different
symbols, Figs. 2 and 3 also show the T--$\rho_{0.5}$ correlations.

In general, the degree of correlation between B/D and $\rho_{\sigma}$ appears
to be similar to that between T and $\rho_{\sigma}$; thus, the partial
correlation coefficient $r$ turns out to be consistently not
significant, except for the KWO' sample, where the B/D--$\rho_{\sigma}$ partial
correlations (for all four $\sigma$-values) turn out to be weakly significant
(on average, at the $\sim$95\% significance level).

We conclude that there is no sure evidence of a true B/D--local
density correlation cleared of spurious secondary dependences related to the
morphology--density relation.

\subsection{The B/D--$\rho_{\sigma}$ and T--$\rho_{\sigma}$ correlations for
combined samples}

We try to build a large sample, combining the SdV, KWO, KWO' and K samples. To
do this, first we multiply all K, KWO, KWO' B/D values by suitable conversion
factors; these, for each morphological type interval, are taken to be the
ratios of the medians of the K, KWO, KWO' values of B/D with respect to the
corresponding medians of the SdV values (see the medians tabulated in Table 1).
For galaxies which are common to two of the three samples, we chose the B/D
values given by SdV and K, in this order of preference. In this manner, for
each
morphological type interval, we construct K, KWO, KWO' B/D distributions, which
by definition have the same medians as the respective SdV distributions,
although they may still have a different shape. Then, we check that the K
normalized B/D distributions are not statistically different from the
corresponding SdV distributions (for each type interval and for all types
together) by applying the classical Kolmogorov-Smirnov test (e.g., Hoel, 1971),
the Rank-Sum test (e.g., Hoel, 1971), and the Mann-Whitney U-test (e.g.,
Kendall \& Stuart, 1979). The same holds true for the comparison between the
KWO, KWO' normalized and SdV distributions. But we realize that the KWO and K
normalized distributions are statistically indistinguishable only for earlier
types (T$<$3), whereas the inclusion of later type objects makes the two
distributions different (e.g., at the 94.3\%, 97.3\%, 98.3\% levels for the
respective intervals T$<$4, T$<$5, T$<$6, according to the Kolmogorov-Smirnov
test). Analogously, as objects of later types are included , the KWO' and K
normalized distributions become different (e.g., at the 96.0\%, 93.4\%, 90.3\%
levels for T$<$4, T$<$5, T$<$6). Therefore, deeming it unreasonable to combine
samples K and KWO (or KWO'), we deal with the combined samples SdV+K, SdV+KWO,
and SdV+KWO' only.

Table 3 presents the correlation analysis of some combined
samples. With respect to the previous analysis of individual samples, combined
samples generally show more significant B/D--$\rho_{\sigma}$ and
T--$\rho_{\sigma}$ total correlations (also for spirals) because there are more
objects, but no appreciable B/D--$\rho_{\sigma}$ partial correlations (not even
for combined samples involving KWO'). This confirms that there is no evidence
of a true dependence of B/D on local density, whenever the effect induced by
the morphology--density relation is removed.

In order to enlarge our samples with the inclusion of non--NBG galaxies, we
provide a further, rough characterization of environmental density,
assigning a parameter ENV to each NBG and non-NBG galaxy. ENV is defined as an
integer which grows with the increasing probability of being a member of the
Virgo cluster. First, we simply assign ENV=0 to non-Virgo objects and ENV=1 to
Virgo members (according to Binngeli, Sandage \& Tammann, 1985). Then, we
reassign ENV=1 to objects located in the peripheral regions of the Virgo
cluster (like the Virgo southern extension and the Virgo clouds W, W', M), and
ENV=2 to members of the main body of the Virgo clusters (Virgo subclusters A
and B) (see also Binggeli, Popescu \& Tammann (1993) for membership assignments
in the Virgo region.). In substantial agreement with the outcomes of our
previous analyses, we can state that there is no clear evidence of a true
B/D--local density correlation in homogeneous and combined samples, whenever
we remove the effect induced by the morphology--density relation.

\subsection{The B/D--$D_v$ and T--$D_v$ correlations}

In a search for large-scale environmental effects on B/D, an interesting
quantity is the galaxy spatial distance $D_v$ (in Mpc) from the center of the
Virgo cluster. As already done in Monaco et al. (1994), we use $D_v$ to explore
effects on large scale in the Local Supercluster. Note that $D_v$, being a
three-dimensional distance, is meaningful mainly outside the Virgo cluster. We
subject our samples to analysis of the B/D--$D_v$ total correlations and
(T-independent) partial correlations. Table 4 presents the results for the most
interesting cases. We detect appreciable T--$D_v$ correlations in wide samples
(S0+S objects of the SdV and KWO samples, K spirals, and combined samples which
involve these data sets; earlier types are typically closer to the Virgo
cluster center (as is expected from the morphology--density relation). This
correlation gives rise to appreciable B/D--$D_v$ total correlations; however,
the partial correlation coefficients are, in general, not significant. Fig. 4
shows the B/D--$D_v$ correlation for the S0+S objects of the KWO sample. Since
various morphological type intervals are denoted by different symbols, Fig. 4
illustrates also the T--$D_v$ correlation. Both correlations are essentially
due to Virgo galaxies; this means that the observed effect is essentially
present on a scale of a couple of Mpc, in agreement with the general
morphology--clustercentric relation (Whitmore et al., 1993). We note that in
the Virgo cluster region the B/D--$D_v$ correlation, while giving essentially
the same information as the B/D--$\rho_{\sigma}$ one, is typically weaker. On
the other hand, no effect is observed on larger scales.

{}From this analysis we conclude that there is no unambiguous evidence that B/D
is related to the distance from the Virgo cluster, irrespective of the tendency
induced by the morphology--density relation.

\subsection{The $M_{bu}$--$\rho_{\sigma}$ and $M_d$--$\rho_{\sigma}$
correlations}

We wish to check whether the absence of significant B/D--$\rho_{\sigma}$
correlations is simply due to the lack of both bulge--absolute magnitude
($M_{bu}$--$\rho_{\sigma}$)  and disk--absolute magnitude
($M_d$--$\rho_{\sigma}$) correlations rather than to a suitable combination of
these two correlations (as, e.g., a parallel dependence of $M_{bu}$ and $M_d$
on environmental density). A similar argument holds for the B/D--$D_v$ and
B/D--ENV correlations. In order to cast light on this question, we undertake an
analysis of the $M_{bu}$--$\rho_{\sigma}$ and $M_d$--$\rho_{\sigma}$
correlations. We consider only the samples SdV, KWO, KWO', because, for many
objects of the SdV sample and all galaxies of the KWO (KWO') sample, the values
of $M_{bu}$ and $M_d$ can be directly estimated from the tabulated structural
parameters and the adopted distance (not from B/D and galaxy absolute
magnitude, which would amplify the uncertainties on $M_{bu}$ and $M_d$). We
analyze the S0+S, the S objects, and two other type intervals, where the
estimates of $M_{bu}$ and $M_d$ are respectively expected to be most
reliable, namely lenticulars and early-type spirals (up to T=3) for $M_{bu}$
and mid/late-type spirals (T$>$2) for $M_d$. We include also objects with
unknown B/D, but known $M_{bu}$ or $M_d$. We construct the KWO' sets of
$M_{bu}$ and $M_d$ data by eliminating from the KWO sample galaxies with
low-quality spheroids and disks (as explained in the previous section). The
$M_{bu}$ and $M_d$ values of the SdV sample are corrected for Galactic and
internal absorption (according to the precepts described by SdV). The $M_{bu}$
and $M_d$ values of the KWO and KWO' samples are uncorrected for these effects;
however, all objects lie in sky areas characterized by similar Galactic
absorption and are not strongly inclined.

Table 5 reports some results relative to the $M_{bu}$--$\rho_{\sigma}$,
$M_d$--$\rho_{\sigma}$ correlations for S0+S objects. The partial
$M_{bu}$--$\rho_{\sigma}$ correlations (at fixed T) are never significant,
although the total ones are significant in some cases (for the S0+S objects of
the SdV and KWO samples). In other words, in the latter cases a dependence of
$M_{bu}$ on environmental density is induced by the morphology--density
relation as well as by the pronounced dimming of bulges towards later types
within the SdV and KWO samples. No total/partial $M_{bu}$--$\rho_{\sigma}$
correlations are observed within the other type intervals.

We find no total or partial $M_d$--$\rho_{\sigma}$ correlations within the SdV
sample. We see an appreciable effect in the S0+S objects of the KWO and KWO'
samples at a relatively large scale (mainly for the parameter $\rho_{2.0}$). No
effects are detected within the other type intervals. In the case of $M_d$,
because of the insignificant or much weaker dimming of the disks with respect
to the bulges towards later types in the samples considered, the
morphology--density relation does not appreciably affect the total
$M_d$--$\rho_{\sigma}$ correlations; i.e., it does not give rise to appreciable
negative total $M_d$--$\rho_{\sigma}$ correlations. We have verified that the
opposite (i.e. positive) effect is induced by the observation
selection of the galaxy sample. Specifically, within a magnitude-limited galaxy
sample which covers mostly the Virgo sky area, the galaxies which lie behind
the main Virgo cluster concentration, in low-density zones, tend to be, on
average, more luminous than their nearby (Virgo cluster member) counterparts;
they will also have, on average, brighter disks. As a matter of fact, we have
checked that all partial correlations disappear when we restrict the samples to
objects lying at a distance smaller than 25 Mpc, which minimizes the effects of
observational selection (see Table 5).

We have verified that the correlations between $M_{bu}$ ($M_d$) and
the other environmental indices $D_v$ and ENV yield consistent results.

To sum up, we find different behaviours of $M_{bu}$ and $M_d$ with the
various environmental indices. The basic difference is to be ascribed to the
influence of the morphology--density relation. In samples covering a wide
morphological interval, this relation alone tends to create a positive bulge
luminosity--density relation, which we can detect in several cases where it
is not masked by the opposite observational selection effect.
The latter effect is then directly reflected in the observed disk
luminosity--density relation, which is indeed little influenced by the
morphology--density relation.

In conclusion, we have verified that all the observed correlations which
involve $M_{bu}$ and $M_d$ are fully explainable as the result of
observational selection, coupled with the effect of the morphology--density
relation. Hence, there is no reason to invoke a primary (T-independent)
influence of the environment on the bulge (or disk) luminosity in order to
explain the observed correlations.

\subsection{The B/D--$M_B$ correlation}

Finally, we investigate whether B/D is related to the galaxy absolute magnitude
$M_B$, which is derived from the adopted distance and the corrected total blue
apparent magnitude $B_{T}$ (tabulated in RC3). To this end, we examine the
B/D--$M_B$ total correlations and (T-independent) partial correlations for all
homogeneous samples and all combined samples. In this case, we also consider
the type interval T$>$2, where a strong B/D--$M_B$ correlation was claimed by
Solanes et al. (1989). Table 6 presents some results for the widest samples
(SdV, K, KWO) and for relevant combined samples. A selection effect of galaxy
samples is responsible for the negative T--$M_B$ correlations which are often
observed for S0+S objects: S0 are generally intrinsically fainter than
early-type spirals in our samples. Moreover, the fact that late-type spirals
are typically less luminous than early-type ones gives rise to the positive
T--$M_B$ correlations which are often detected for S objects. The B/D--$M_B$
total correlation is mostly not significant, but in some cases the T--$M_B$
correlations induce significant B/D--$M_B$ total correlations (of opposite
sign). In any case, the B/D--$M_B$ partial correlation is never significant for
homogeneous and combined samples of NBG objects (for all type intervals
considered).

We repeat this correlation analysis by including in each sample also objects
(with known $B_{T}$ and redshift) which are not listed in NBG. For non-Virgo
galaxies we simply adopt redshift-distances; for the sake of consistency with
the NBG galaxy distances, members of the main body of the Virgo cluster (Virgo
subclusters A and B) and of the Virgo cloud W' are given a distance of 16.8
Mpc, while members of the Virgo clouds W and M are given a distance of 35.1
Mpc. For samples enlarged to non-NBG galaxies, there is generally no B/D--$M_B$
relation, although in some samples (the S0+S objects of the K, SdV+KWO, SdV+K
samples) a negative T--$M_B$ correlation induces a marginal B/D--$M_B$ partial
correlation (at the $<$96\% level only). If we repeat the analysis for the
subsamples of S0 and S objects, also this weak partial correlation typically
vanishes.

Therefore, it is safe to conclude that there is no good evidence that
B/D is related to $M_B$, for a given morphological type. This conclusion is
statistically well-founded especially for mid- and late-type spirals (T$>$2).)

\section{Discussion and Conclusions}

Considering the role of the known morphology--density relation,
from a statistical analysis of the best available B/D data sets we draw
the following major conclusions:

1) The tendency of galaxies to have greater B/D with increasing local density
(within the widest samples) is a reflection of the morphology--density
relation, since this tendency appears to be simply due to the average decline
of B/D towards later types. But this was not obvious a priori. To be more
specific, when we remove the effect induced by the morphology--density
relation, we find no clear evidence of a B/D dependence on the local density,
at either large ($\sim$a few Mpc) or small ($\sim$a few tenths of Mpc) scales
in the very nearby ($z<0.01$) volume of the universe (The Local Supercluster).

2) Unlike previous claims, for a given morphological type, B/D is found to
be substantially unrelated to the galaxy absolute magnitude (and, hence,
probably to the galaxy total mass).

3) There is no need to hypothesize a primary dependence of bulge and disk
luminosities on environmental density, as we note that the morphology--density
relation alone tends to give rise to a greater proportion of bright bulges in
denser regions (which explains some earlier claims). As a matter of fact, the
growth of B/D towards earlier types appears to be mostly due to a pronounced
brightening of the bulge rather than to a dimming of the disk. Incidentally,
Solanes et al.'s (1989) study of Dressler's sample yielded the opposite
conclusion on the latter point. In other words, our study emphasizes that the
morphology--density relation is likely to be coupled with a density segregation
of bulge luminosity  rather than with a segregation of disk luminosity. This
point was regarded as unclear in Oemler's (1992) review.

Let us now discuss to what extent our findings differ from previous results.

Point 1 confirms the absence of primary environmental effects on B/D at small
scales (of a few tenths of Mpc), in agreement with the conclusions reached by
Solanes et al. (1989). On the basis of a different data set and using a
two-dimensional definition of the local galaxy density, these authors clarified
the (previously not well understood) role played by the morphology--density
relation. Furthermore, we report the absence of (as yet unexplored) effects on
B/D also at large scales (of $\sim$a few Mpc). Interestingly, this latter
finding does not support a large-scale density segregation of bulge masses (at
fixed morphological type), which has been recently hypothesized in order to try
to explain the observed large-scale segregation of local low-luminosity active
galactic nuclei hosted in luminous early-type spirals  (see Monaco et al.,
1994).

Point 2 disagrees with previous controversial claims mentioned in the
introduction, which are likely to be  affected by the following problems. In
wide samples of mixed morphological types, a (spurious) B/D--$M_B$ correlation
may simply arise from differences in the optical luminosity functions of
objects of different morphological types. These differences are especially
marked between bright ellipticals, lenticulars, early spirals and later types
(late spirals and irregulars) (see, e.g., Binggeli, Sandage \& Tammann, 1988;
Efstathiou, Ellis \& Peterson, 1988; Santiago \& Strauss, 1992). This may
account for the claim by Schechter \& Dressler (1987), who did not use a
morphological type subdivision of the objects, if in their sample early-type
objects are typically more luminous than late-type ones.

But even if this subdivision is applied to Dressler's (1980a)
sample (as was done by Solanes et al, 1989), two main problems remain: $i)$ the
presence of a number of unresolved bulges, for which Solanes et al. (1989)
assumed a blue apparent magnitude of $m=19.5$; $ii)$ a severe field
contamination among cluster members; in other words, a number of late-type
spirals with unknown redshift could be low-luminosity foreground objects rather
than bright cluster members. These problems make the B/D--$M_B$
positive correlation claimed by the Spanish authors very doubtful.

In agreement with our view (point 3), Solanes et al. (1989) found no evidence
of direct environmental effects on the bulge (or disk) luminosity by using
an indirect approach. They showed that if one takes galaxy samples which are
bulge-limited instead of total magnitude-limited, the luminosity--density
relation (claimed in previous studies) disappears, essentially because in this
way only early-type galaxies (spanning a limited type interval) are left in the
sample. This is tantamount to eliminating much of the bias induced by the
morphology--density relation.

However, their interpretation of the effect of morphological segregation with
density is radically different. From an inspection of
Dressler's sample, by indirect reasoning they argued that a
decline in disk luminosities towards earlier types is the basis of the
observed dependence of B/D on morphology. Consequently, in their view, the
seeming tendency of B/D to increase with increasing local density is attributed
to a decrease in disk luminosities rather than to an increase in bulge
luminosities. The latter view is, instead, more correct even for Dressler's
sample, as is readily proved by the plots presented by Lake \& Carlberg
(1988), who showed that B/D is much more closely related to the bulge
than to the disk luminosities.

This question is important in the context of theories of galaxy formation and
evolution because the mechanisms which can alter disks or bulges
(see, e.g., the reviews by Evrard (1992) and Mamon (1993)) are generally
different. A scenario in which bulges appear to be more affected by the
environment than disks tends to favour processes occurring
at galaxy formation (or at the beginnning of galaxy life) --- rather
than late evolutionary processes --- as being the major ones
responsible for the morphological segregation of disk galaxies.
The fact that primordial star formation rate and merger events are expected
to proceed more rapidly in denser environments may be sufficient to
account, at least qualitatively, for the observed dependence of morphology
and bulge luminosity on environmental density (see, e.g., Larson,
1992). A basic structural property like B/D is probably much less affected
by continuing environmental effects (i.e. by the late evolutionary history of
a galaxy) than other galaxy properties such as bars, arm classes, rings,
tails, bridges, and gas content. Advances in the numerical simulation
of galaxy formation in a cosmological context (e.g., along the lines followed
by Katz (1992); Steinmetz \& M\"uller (1993); Navarro \& White (1994))
promise to yield quantitative predictions on the relative proportion of bulges
and disks in galaxies.

Our results imply no appreciable total luminosity segregation with density for
a given morphological type, and a slight luminosity segregation for samples of
galaxies spanning a rather wide interval of morphologies (e.g., for generic
samples of disk or spiral galaxies), at least within our nearby universe. In
recent years, there has been a growing number of studies on redshift
catalogues,
which provide evidence of some total luminosity segregation (weaker than
morphology segregation) for luminous galaxies (e.g., Iovino et al., 1993;
Dom\'inguez-Tenreiro, G\'omez-Flechoso \& Mart\'inez, 1994) or at least for
early-type luminous galaxies, whenever corrections for Galactic and internal
absorptions are applied to galaxy magnitudes (e.g., Hasegawa \& Umemura,
1993). There is also some evidence of diameter segregation for galaxies
with large (face-on) diameters (e.g., Fang \& Zou, 1994; Campos,
Dom\'inguez-Tenreiro \& Yepes, G., 1994).

Wide galaxy samples with accurate, homogeneous B/D decomposition
would be very valuable for further investigations into the presence of primary,
subtle environmental effects, which are undetectable on the basis of the
available data.

\bigskip
\bigskip

We thank Y. C. Andredakis and R. H. Sanders, who kindly sent us some
results of their study of photometric decomposition.

One of us (F. L. T.) acknowledges the kind hospitality provided by
the Department of Astronomy of Trieste University, SISSA, and the
International Center for Theoretical Physics during her stay in Trieste.

This work has been partially supported by the Italian Research Council
(CNR-GNA) and by the Ministry of University, Scientific and
Technological Research (MURST).

\newpage
{\bf Table 1:} means, medians, and standard deviations of the
B/D-distributions.

\bigskip
\bigskip

\begin{center} \small
\begin{tabular}{l|c|c|c|c|c|c}
& & & & & &\\
Sample & N & T & mean & s.d. & median & 90\% c. i.\\
& & & & & &\\
\hline\hline & & & & & &\\
SdV   &    28  &  -3,-2,-1 &   1.80 &  1.36 &   1.58  &   1.06 -- 1.71\\
SdV   &     7  &    0,1    &   1.63  &  1.83 &   1.20  &   0.28 -- 1.45\\
SdV   &    15  &      2    &   0.65 &  0.47 &   0.56  &   0.23 -- 0.86\\
SdV   &    15  &      3    &   0.46 &  0.49 &   0.26 &    0.18 -- 0.64\\
SdV   &    13  &      4    &   0.28 &  0.22 &   0.24 &    0.13 -- 0.30\\
SdV   &     9   &     5    &   0.14 &  0.10 &   0.10 &    0.06 -- 0.18\\
SdV   &    10  &     6,7   &   0.05 &  0.03 &   0.04 &    0.02 -- 0.09\\
K     &    16  &  -3,-2,-1 &   3.36 &  2.94 &   1.70 &    0.89 -- 5.25\\
K     &    14  &     0,1   &   2.16 &  2.86 &   0.88  &   0.54 -- 2.33\\
K     &    15  &      2    &   0.81 &  0.90 &   0.52  &   0.18 -- 0.82\\
K     &    19  &      3    &   0.28 &  0.31 &   0.18  &   0.08 -- 0.41\\
K     &    18  &      4    &   0.17 &  0.23 &   0.09  &   0.05 -- 0.18\\
K     &    23  &      5    &   0.12 &  0.17 &   0.03  &   0.02 -- 0.11\\
K     &     5  &     6,7   &   0.51 &  1.02 &   0.02  &   0.00 -- 2.33\\
KWO   &    50  &  -3,-2,-1 &   2.61 &  2.11 &   2.29  &   1.45 -- 2.88\\
KWO   &    23  &     0,1   &   1.60 &  1.87 &   0.91  &   0.58 -- 0.91\\
KWO   &     6  &      2    &   1.80 &  1.06 &   1.63  &   0.00 -- 3.63\\
KWO   &     8  &      3    &   0.47 &  0.28 &   0.41  &   0.15 -- 0.72\\
KWO   &    11  &      4    &   0.35 &  0.18 &   0.36  &   0.12 -- 0.58\\
KWO   &    11  &      5    &   0.37 &  0.50 &   0.15  &   0.15 -- 0.46\\
KWO   &    20  &    $>$5    &   0.86 &  1.02 &   0.46  &   0.15 -- 0.72\\
KWO'  &    23  &  -3,-2,-1 &   2.42 &  2.10 &   1.82  &   1.45 -- 2.29\\
KWO'  &    11  &     0,1   &   0.89 &  0.76 &   0.72  &   0.15 -- 1.82\\
KWO'  &     3  &      2    &   1.15 &  0.59 &   0.91  &   0.00 -- 1.82\\
KWO'  &     6  &      3    &   0.45 &  0.31 &   0.23  &   0.12 -- 0.36\\
KWO'  &     7  &      4    &   0.25 &  0.13 &   0.23  &   0.12 -- 0.36\\
KWO'  &     8  &      5    &   0.41 &  0.58 &   0.16  &   0.12 -- 0.46\\
KWO'  &    15  &    $>$5    &   0.48 &  0.52 &   0.18  &   0.14 -- 0.46\\
AS1   &    11  &     2,3   &   0.67 &  1.56 &   0.07  &   0.04 -- 0.45\\
AS1   &     9  &      4    &   0.59 &  0.86 &   0.07  &   0.02 -- 0.77\\
AS1   &    12  &    5 -- 7  &   0.30 &  0.42 &   0.04  &  0.003 -- 0.74\\
AS2   &    11  &     2,3   &   0.05 &  0.03 &   0.05  &   0.02 -- 0.08\\
AS2   &     9  &      4    &   0.10 &  0.12 &   0.03  &   0.01 -- 0.15\\
AS2   &    12  &    5 -- 7  &   0.26 &  0.78 &  0.03 &   0.004 -- 0.04\\
\end{tabular} \normalsize
\end{center}

\newpage

{\bf Table 2:} the B/D--$\rho_{\sigma}$, T--$\rho_{\sigma}$ correlations
for the homogeneous samples.

\bigskip
\bigskip

\begin{center} \small
\begin{tabular}{l|c|c|c|c|c}
& & & & &\\
Sample   &     N   &  Variables   &  $r_s$ & $r_k$  &  $r$\\
& & & & &\\
\hline\hline & & & & &\\
SdV (S0+S) &94 & B/D-$\rho_{0.5}$& 0.263 (99.5) &0.185 (99.6)& 0.035 ($<$90.)\\
SdV (S0+S) &  94 & T-$\rho_{0.5}$ & -0.358 (99.98)& -0.253 (99.99)&    \\
KWO (S0+S)&100&B/D-$\rho_{0.5}$&0.249 (99.4)&0.173 (99.5)&0.075 ($<$90.)\\
KWO (S0+S)&100&T-$\rho_{0.5}$& -0.324 (99.95)& -0.231 (99.97)&   \\
KWO' (S0+S)&56&B/D-$\rho_{0.5}$& 0.337 (99.4)&0.230 (99.4)&0.154 (96.8)\\
KWO' (S0+S)&56&T-$\rho_{0.5}$& -0.299 (98.7)&-0.222 (99.2)&   \\
K (S0+S)&59&B/D-$\rho_{0.5}$& 0.055 ($<$90.)&0.044 ($<$90.)&  --\\
K (S0+S)&59&T-$\rho_{0.5}$& -0.021 ($<$90.)&-0.018 ($<$90.)&   \\
\end{tabular} \normalsize
\end{center}

\bigskip
\bigskip

{\bf Table 3:} the B/D--$\rho_{\sigma}$, T--$\rho_{\sigma}$ correlations
for the combined samples.

\bigskip
\bigskip

\begin{center} \small
\begin{tabular}{l|c|c|c|c|c}
& & & & &\\
Sample &         N &  Variables &     $r_s$   &     $r_k$   &   $r$\\
& & & & &\\
\hline\hline & & & & &\\
SdV+K (S0+S)&141&B/D-$\rho_{0.5}$&0.233 (99.7)&0.161 (99.8)&0.067 ($<$90.)\\
SdV+K (S0+S)&141&T-$\rho_{0.5}$& -0.276 (99.95)&-0.195 (99.97)&    \\
SdV+KWO (S0+S)&175&B/D-$\rho_{0.5}$&0.256 (99.97)&0.180 (99.98)&0.062
($<$90.)\\
SdV+KWO (S0+S)&175&T-$\rho_{0.5}$&-0.309 ($>$99.99)&-0.216 ($>$99.99)&  \\
SdV+KWO' (S0+S)&139&B/D-$\rho_{0.5}$&0.230 (99.7)&0.159 (99.7)&0.066 ($<$90.)\\
SdV+KWO' (S0+S)&139&T-$\rho_{0.5}$&-0.251 (99.9)&-0.178 (99.9)&   \\
\end{tabular} \normalsize
\end{center}

\newpage

{\bf Table 4:} the B/D-$D_v$, T-$D_v$ correlations.

\bigskip
\bigskip

\begin{center} \small
\begin{tabular}{l|c|c|c|c|c}
& & & & &\\
Sample & N &  Variables  & $r_s$   &  $r_k$   & $r$\\
& & & & &\\
\hline\hline & & & & &\\
SdV (S0+S)&94& B/D-$D_v$ & -0.136 (90.4)&-0.091 (90.4)& -0.020 ($<$90.)\\
SdV (S0+S)&94&T-$D_v$&0.175 (95.4)&0.121 (95.8)&    \\
KWO (S0+S)&100&B/D-$D_v$ & -0.198 (97.6)& -0.139 (98.0)&-0.051 ($<$90.)\\
KWO (S0+S)&100&T-$D_v$ &0.291 (99.8) &0.202 (99.9)&      \\
KWO' (S0+S)&56& B/D-$D_v$&-0.103 ($<$90.)&-0.071 ($<$90.)&  --\\
KWO' (S0+S)&56&T-$D_v$&0.152 ($<$90.)&0.102 ($<$90.)&    \\
K (S0+S)&59& B/D-$D_v$& -0.091 ($<$90.)& -0.059 ($<$90.)& --\\
K (S0+S)&59&T-$D_v$&0.082 ($<$90.)&0.049 ($<$90.)&   \\
SdV+K (S0+S)&141 &B/D-$D_v$& -0.150 (96.2)&-0.101 (96.3)& -0.042 ($<$90.)  \\
SdV+K (S0+S)&141&T-$D_v$&0.176 (98.2)& 0.121 (98.4)&    \\
SdV+KWO (S0+S)& 175& B/D-$D_v$&-0.163 (98.5)&-0.112 (98.6)& -0.027 ($<$90.)\\
SdV+KWO (S0+S)& 175&T-$D_v$& 0.212 (99.8)&0.149 (99.8)&      \\
SdV+KWO' (S0+S)&139&B/D-$D_v$&-0.068 ($<$90.)& -0.046 ($<$90.)& --\\
SdV+KWO' (S0+S)&139&T-$D_v$& 0.090 ($<$90.)&0.061 ($<$90.)&     \\
\end{tabular} \normalsize
\end{center}

\newpage

{\bf Table 5:} the $M_{bu}$--$\rho_{\sigma}$, $M_d$--$\rho_{\sigma}$
correlations.

\bigskip
\bigskip

\begin{center} \small
\begin{tabular}{l|c|c|c|c|c}
& & & & &\\
Sample &        N  &   Variables   &  $r_s$    &     $r_k$  &   $r$\\
& & & & &\\
\hline\hline & & & & &\\
SdV (S0+S)&94&$M_{bu}-\rho_{0.5}$&-0.176 (95.5)&-0.132 (97.0)&-0.049 ($<$90.)\\
SdV (S0+S)&94&$M_{bu}-\rho_{1.0}$&-0.218 (98.3)&-0.157 (98.8)&-0.081 ($<$90.)\\
SdV (S0+S)&94&$M_d-\rho_{0.5}$&0.123 ($<$90.)&0.080 ($<$90.)&0.022 ($<$90.)\\
SdV (S0+S)&94&$M_d-\rho_{1.0}$&0.072 ($<$90.)&0.040 ($<$90.)&-0.019 ($<$90.)\\
SdV (S0+S)&94&$M_d-\rho_{2.0}$&-0.009 ($<$90.)&-0.002 ($<$90.)&-0.064
($<$90.)\\
KWO (S0+S)&100&$M_{bu}-\rho_{0.5}$&-0.175 (95.9)&-0.122 (96.4)&-0.037
($<$90.)\\
KWO (S0+S)&100&$M_{bu}-\rho_{1.0}$&-0.122 ($<$90.)&-0.077 ($<$90.)&0.021
 ($<$90.)\\
KWO (S0+S)&106&$M_d-\rho_{0.5}$&0.140 (92.4)&0.087 (90.7)&0.076 ($<$90.)\\
KWO (S0+S)&106&$M_d-\rho_{1.0}$&0.186 (97.2)&0.125 (97.2)&0.115 (96.3)\\
KWO (S0+S)&106&$M_d-\rho_{2.0}$&0.241 (99.4)&0.160 (99.3)&0.151 (98.8)\\
KWO (S0+S) D$<$25Mpc&93&$M_d-\rho_{2.0}$&0.085 ($<$90.)&0.059 ($<$90.)&0.052
  ($<$90.)\\
KWO' (S0+S)&56&$M_{bu}-\rho_{0.5}$&-0.175 (90.1)&-0.129 (91.9)&-0.057
($<$90.)\\
KWO' (S0+S)&56&$M_{bu}-\rho_{1.0}$&-0.129 ($<$90.)&-0.078 ($<$90.)&-0.007
($<$90.)\\
KWO' (S0+S)&91&$M_d-\rho_{0.5}$&0.091 ($<$90.)&0.056 ($<$90.)&0.057 ($<$90.)\\
KWO' (S0+S)&91&$M_d-\rho_{1.0}$& 0.139 (90.5)&0.097 (91.4)&0.100 (93.2)\\
KWO' (S0+S)&91&$M_d-\rho_{2.0}$&0.194 (96.7)&0.129 (96.5)&0.131 (97.0)\\
KWO' (S0+S) D$<$25 Mpc&78&$M_d-\rho_{2.0}$&0.053 ($<$90.)&0.038 ($<$90.)&0.041
 ($<$90.)\\
\end{tabular} \normalsize
\end{center}

\bigskip
\newpage

{\bf Table 6:} the B/D-$M_B$, T-$M_B$ correlations.

\bigskip
\bigskip

\begin{center} \small
\begin{tabular}{l|c|c|c|c|c}
& & & & &\\
Sample &          N  &   Variables &    $r_s$  &  $r_k$    &  $r$\\
& & & & &\\
\hline\hline & & & & &\\
SdV (S0+S)&94&B/D-$M_B$&0.076 ($<$90.)&0.053 ($<$90.)&-0.079 ($<$90.)\\
SdV (S0+S)&94&T-$M_B$&-0.230 (98.7)&-0.179 (99.5)&   \\
SdV (S) &67 &B/D-$M_B$ & -0.102 ($<$90.)& -0.069 ($<$90.)& --\\
SdV (S) &67 &T-$M_B$ & 0.015 ($<$90.) & -0.003 ($<$90.)&  \\
KWO (S0+S)&100&B/D-$M_B$& 0.110 ($<$90.)&0.082 (90.8)& --\\
KWO (S0+S)&100&T-$M_B$&-0.050 ($<$90.)& -0.048 ($<$90.)&     \\
KWO (S) &59 &B/D-$M_B$& 0.083 ($<$90.)& 0.055 ($<$90.)&0.102 ($<$90.)\\
KWO (S) &59 &T-$M_B$& 0.211 (94.6)&0.152 (95.5) &  \\
KWO' (S0+S)&56&B/D-$M_B$&0.027 ($<$90.)&0.016 ($<$90.)&  --\\
KWO' (S0+S)&56&T-$M_B$& -0.041 ($<$90.)&-0.056 ($<$90.)&     \\
KWO' (S)& 38&B/D-$M_B$& -0.043 ($<$90.)& -0.035 ($<$90.)&-0.005 ($<$90.)\\
KWO' (S)& 38&T-$M_B$& 0.295 (96.4)& 0.205 (96.5)&    \\
K (S0+S)&59&B/D-$M_B$&-0.126 ($<$90.)&-0.092 ($<$90.)&   --\\
K (S0+S)&59&T-$M_B$&-0.111 ($<$90.)&-0.066 ($<$90.)&     \\
K (S)&51&B/D-$M_B$&-0.115 ($<$90.)&-0.084 ($<$90.)&    \\
K (S)&51&T-$M_B$&-0.177 ($<$90.)&-0.113 ($<$90.)&      \\
SdV+K (S0+S)&141&B/D-$M_B$&0.019 ($<$90.)&0.021 ($<$90.)&-0.069 (90.3)\\
SdV+K (S0+S)&141&T-$M_B$&-0.195 (99.0)&-0.143 (99.4)&    \\
SdV+K (S)&109&B/D-$M_B$&-0.135 (91.8)&-0.084 (90.1)&   --\\
SdV+K (S)&109&T-$M_B$&-0.046 ($<$90.)&-0.036 ($<$90.)&    \\
SdV+KWO (S0+S)&175&B/D-$M_B$&-0.002 ($<$90.)&0.002 ($<$90.)&-0.072 (91.5)\\
SdV+KWO (S0+S)&175&T-$M_B$&-0.115 (93.5)&-0.095 (96.9)&    \\
SdV+KWO (S)&117&B/D-$M_B$& -0.173 (96.9)& -0.115 (96.7)&-0.056 ($<$90.)\\
SdV+KWO (S)&117&T-$M_B$& 0.171 (96.7)&0.115 (96.7)&    \\
SdV+KWO' (S0+S)&139& B/D-$M_B$&-0.024 ($<$90.)&-0.017 ($<$90.)&  --\\
SdV+KWO' (S0+S)&139&T-$M_B$&-0.071 ($<$90.)&-0.068 ($<$90.)&     \\
SdV+KWO' (S)&98&B/D-$M_B$&-0.158 (94.0)&-0.108 (91.3)&-0.035 ($<$90.)\\
SdV+KWO' (S)&98&T-$M_B$& 0.220 (98.5)& 0.142 (98.1)&    \\
\end{tabular}
\normalsize
\end{center}

\newpage
\section*{References}

\ref Andredakis, Y. C. \& Sanders, R. H., 1994, MNRAS, 267, 283. (AS)

\ref Binggeli, B., Popescu, C. C. \& Tammann, G. A., 1993, A\&AS, 98, 275.

\ref Binggeli, B., Sandage, A. \& Tammann, G. A., 1985, AJ, 90, 1681.

\ref Binggeli, B., Sandage, A. \& Tammann, G. A., 1988, ARA\&A, 26, 509.

\ref Campos, A., Dom\'inguez-Tenreiro, R.  \& Yepes, G., 1994, ApJ, 436, 565.

\ref Capaccioli, M. \& Caon, N., 1992, in "Morphological and Physical
Classification of Galaxies", eds. G. Longo, M. Capaccioli \& G. Busarello
(Dordrecht: Kluwer Academic Publ.), 99.

\ref de Vaucouleurs, G., de Vaucouleurs, A., Corwin, H. G., Buta, R. J.,
Paturel, G. \& Fouqu\'e, P., 1991, "Third Reference Catalog of Bright
Galaxies" (New York: Springer Verlag) (RC3).

\ref Dom\'inguez-Tenreiro, R., G\'omez-Flechoso, M. A. \& Mart\'inez, V. J.,
1994, ApJ, 424, 42.

\ref Dressler, A., 1980a, ApJS, 42,565.

\ref Dressler, A., 1980b, ApJ, 236, 351.

\ref Efron, B., 1979, SIAM Rev., 21, 460.

\ref Efron, B. \& Tibshirani, R., 1985, Behaviormetrika, 17, 1.

\ref Efstathiou, G., Ellis, R. S. \& Peterson, B. A., 1988, MNRAS, 232, 431.

\ref Evrard, A. E., 1992, in "Physics of Nearby Galaxies, Nature or Nurture?",
eds. T. X. Thuan, C. Balkowski \& J. Tran Thanh Van (Gif-sur-Yvette:
Editions Frontieres), 375.

\ref Fang, F. \& Zou, Z.--L., 1994, ApJ, 421, 9.

\ref Giuricin, G., Mardirossian, F., Mezzetti, M. \& Monaco, P., 1993,
ApJ, 407, 22.

\ref Giuricin, G., Monaco, P., Mardirossian, F., Mezzetti, M., 1994, ApJ,
425, 450.

\ref Giuricin, G., Mardirossian, F., Mezzetti, M., Pisani, A. \& Ramella, M.,
1988, A\&A, 192, 95.

\ref Hasegawa, T. \& Umemura, M., 1993, MNRAS, 263, 191.

\ref Hoel, P. G., 1971, Introduction to Mathematical Statistics (New York:
Wiley).

\ref Iovino, A., Giovanelli, R., Haynes, M., Chincarini, G. \& Guzzo, L., 1993,
MNRAS, 265, 21.

\ref Katz, N., 1992, ApJ, 391, 502.

\ref Kendall, M. \& Stuart, A., 1977, The Advanced Theory of Statistics
(London:
     Griffin \& Co).

\ref Kent, S., 1985, ApJS, 59, 115. (K85)

\ref Kent, S., 1986, AJ, 91, 1301. (K86)

\ref Kent, S., 1988, AJ, 96, 514. (K88)

\ref Kodaira, K., Watanabe, M. \& Okamura, S., 1986, ApJS, 62, 703. (KWO)

\ref Kormendy, J., 1977, ApJ, 217, 406.

\ref Lake, G. \& Carlberg, R. G., 1988, AJ, 96, 1587.

\ref Larson, R. B., 1992, in "Physics of Nearby Galaxies, Nature or Nurture?",
eds. T. X. Thuan, C. Balkowski \& J. Tran Thanh Van (Gif-sur-Yvette:
Editions Frontieres), 487.

\ref Mamon, G., 1993, in "N-Body Problems and Gravitational Dynamics", eds.
F. Combes \& E. Athanassoula (Paris: Observatoire de Paris), 226.

\ref Monaco, P., Giuricin, G., Mardirossian, F. \& Mezzetti, M., 1994,
ApJ, 436, 576.

\ref Navarro, J. F. \& White, S., 1994, MNRAS, 267, 401

\ref Oemler, A., Jr., 1992, in "Clusters and Superclusters of Galaxies",
eds. A. C. Fabian (Dordrecht: Kluwer Academic Publ.), 29.

\ref Santiago, B. X. \& Strauss, M. A., 1992, ApJ, 387, 9.

\ref Schechter, P. L. \& Dressler, A., 1987, AJ, 94, 563.

\ref Schombert, J. A. \& Bothun, G. D., 1987, AJ, 92, 60.

\ref Siegel, S., 1956, Nonparametric Statistics for the Behavioural Sciences
(New York: McGraw-Hill), 68.

\ref Simien, F., 1988, in "The World of Galaxies", eds. H. G. Corwin, Jr. \&
L. Bottinelli (Springer Verlag), 293.

\ref Simien, F. \& de Vaucouleurs, G., 1986, ApJ, 302, 564. (SdV)

\ref Solanes, J. M., Salvador-Sol\'e, E. \& Sanrom\`a, M., 1989, AJ, 98, 798.

\ref Steinmetz, M. \& M\"uller, 1993, A\&A, 268, 391.

\ref Thuan, T. X. \& Gunn, J. E., 1976, PASP, 88, 543.

\ref Tully, R. B., 1988a, Nearby Galaxies Catalog (Cambridge: Cambridge Univ.
     Press) (NBG).

\ref Tully, R. B., 1988b, AJ, 96, 73.

\ref Tully, R. B. \& Shaya, E. J., 1984, ApJ, 281, 31.

\ref Whitmore, B. C., 1990, in "Clusters of Galaxies", eds. W. R. Oegerl,
M. S. Fitchett \& L. Danly (Cambridge: Cambridge University Press), 139.

\ref Whitmore, B. C., Gilmore, D. M. \& Jones, C., 1993, ApJ, 407, 489.

\newpage
\section*{Figure Captions}
\bigskip

{\bf Fig. 1:} plot of the median values of B/D, together with the
associated 90\% confidence intervals, for different morphological type
intervals (as reported in Table 1). The values relative to the SdV, K,
KWO, and KWO' samples are denoted by circles, crosses, triangles,
and squares, respectively.

{\bf Fig. 2:} plot of the B/D--$\rho_{0.5}$ correlation for the S0+S
objects of the SdV sample. The lenticulars, early--type spirals, and
late--type spirals (T$>$3) are denoted by open circles, crosses, and
dots, respectively.

{\bf Fig. 3:} plot of the B/D--$\rho_{0.5}$ correlation for the S0+S
objects of the KWO sample. Symbols as in Fig. 2

{\bf Fig. 4:} plot of the B/D--$D_v$ correlation for the S0+S objects
of the KWO sample. Symbols as in Fig. 2.

\end{document}